\documentclass{aip-cp}

\usepackage[numbers]{natbib}
\usepackage{rotating}
\usepackage{graphicx}

\def\gtap{\ \raise.3ex\hbox{$>$\kern-.75em\lower1ex\hbox{$\sim$}}\ }
\def\ltap{\ \raise.3ex\hbox{$<$\kern-.75em\lower1ex\hbox{$\sim$}}\ }

\begin{document}

\title{Dynamical coupled-channels approach to electroweak meson
productions on nucleon and deuteron}

\author[aff1,aff2]{Satoshi X. Nakamura\corref{cor1}}

\affil[aff1] {University of Science and Technology of China, Hefei 230026, 
People's Republic of China}
\affil[aff2]{State Key Laboratory of Particle Detection and Electronics (IHEP-USTC), Hefei 230036, People's Republic of China}
\corresp[cor1]{Corresponding author: satoshi@ustc.edu.cn}

\maketitle

 \begin{abstract}
  I overview our recent activity with
the Argonne-Osaka dynamical coupled-channels (DCC) approach that provides a
unified description of various electroweak meson productions on single
 nucleon and nucleus.
First I discuss the DCC model of a single nucleon. 
The DCC model has been developed through a comprehensive analysis of 
$\pi N, \gamma N\to \pi N, \eta N, K\Lambda, K\Sigma$ reaction data.
The model has been further extended
to finite $Q^2$ region by analyzing pion electroproduction data, and
  to neutrino-induced reactions using the PCAC relation.
Next I discuss applications of the
DCC model to electroweak meson productions on the deuteron. We
consider impulse mechanism supplemented by final state interactions (FSI)
due to $NN$ and meson-nucleon rescatterings.
Using this model,
  I discuss FSI corrections needed to extract $\gamma$-neutron 
  reaction observables from $\gamma d\to \pi NN$, and 
  a novel method to extract $\eta N$ scattering length from
  $\gamma d\to \eta pn$.
  I also discuss FSI corrections on the existing neutrino-nucleon pion
  production data that had been
extracted from neutrino-deuteron data. 
 \end{abstract}

\section{Introduction}

Electroweak meson productions on nucleon and deuteron have been
attracting physicists' interests.
This is primarily because the reactions are very useful 
for studying the baryon spectroscopy.
By analyzing data of these processes, we can identify nucleon resonances and
extract their properties such as mass, width, and electromagnetic form factors. 
Combining these properties with outputs from hadron structure models
and Lattice QCD, we can better understand QCD in the nonperturbative
regime. 
Our another interest in the processes is to obtain a basis to study
electroweak meson productions on nuclei. 
Using the obtained single nucleon amplitudes as a basis, we can address
medium modifications on the propagations of mesons and nucleon
resonances in nuclear matter. 
We can also study neutrino-nucleus reactions, understanding of which is 
highly demanded for extracting the neutrino properties from ongoing
neutrino oscillation experiments.

The Argonne(ANL)-Osaka Collaboration~\cite{web} has proved
a dynamical coupled-channels (DCC) approach
very successful in describing 
electroweak meson productions on nucleon and deuteron, 
and in extracting nucleon resonance properties. 
While other common framework such as the K-matrix model~\cite{bg12}
has been also successful in this regard, 
a unique feature in the DCC approach is that it provides reaction
mechanisms and dynamical contents of the nucleon resonances in terms of
hadronic degrees of freedom. 
Also, the DCC model for meson-baryon reactions is compatible well with
multiple scattering theory for many-body system, and thus applications of
amplitudes from the DCC model to nuclear processes can be done
straightforwardly (not necessarily easy though) without introducing
artificial prescriptions. 

This contribution is about reviewing our recent activity with the DCC
approach. 
The former part is on the DCC approach to the single nucleon sector. 
First we discuss the DCC analysis of 
$\pi N, \gamma N\to \pi N, \pi\pi N, \eta N, K\Lambda, K\Sigma$ reactions.
Then the DCC model is extended to finite $Q^2$ region and
neutrino-induced meson productions. 
The latter part is about applications of the DCC model to
electroweak meson productions on the deuteron.
Specifically we discuss three subjects. 
The first one is about extracting neutron-target observables from 
$\gamma d\to \pi NN$. 
The second one is about a novel method of extracting 
$\eta$-nucleon scattering length from
$\gamma d\to \eta np$.
The last one is on 
FSI corrections to neutrino-nucleon cross section data from
neutrino-deuteron experiments.

\section{Dynamical coupled-channels approach to
$\pi N, \gamma N\to \pi N, \pi\pi N, \eta N, K\Lambda, K\Sigma$}

The dynamical coupled-channels model is designed to describe
electroweak meson productions off the nucleon in the resonance region
($W\ltap$ 2~GeV).
It is based on a coupled-channel Lippmann-Schwinger equation that takes
care of couplings among 
$\pi N, \pi\pi N, \eta N, K\Lambda, K\Sigma, \pi\Delta, \rho N, \sigma N$ 
channels to satisfy the unitarity.
The electroweak couplings are considered perturbatively.
The meson-baryon interactions in the model consist of non-resonant
meson-exchange mechanisms and resonant bare $N^*$-excitation
mechanisms. 
By solving the Lippmann-Schwinger equation, the bare $N^*$ states are dressed by
meson clouds to form nucleon resonances. 
The DCC model has been developed through a comprehensive analysis of 
$\pi N, \gamma N\to \pi N, \eta N, K\Lambda, K\Sigma$ reaction data;
the number of data points amounts to be $\sim 27,000$.
The quality of fitting the data achieved with the DCC model can be
found in Refs.~\cite{knls13,knls16}.
The parameters associated with the nucleon resonances such as pole positions
and helicity amplitudes have been successfully extracted from
the DCC model amplitudes~\cite{knls13,knls16}.
All ANL-Osaka DCC analysis results and 
partial wave amplitudes are collected in Ref.~\cite{web}.

\section{Extension of the DCC model to finite $Q^2$ region and
 neutrino reactions}

Ongoing and near-future neutrino oscillation experiments such as T2K~\cite{t2k} and
DUNE~\cite{dune} primarily aim at discovering the CP violation in the lepton sector and
determining the neutrino mass hierarchy.
They need a high-precision neutrino-nucleus pion production model
for this purpose, because neutrino-nucleus reactions 
they utilize to detect neutrinos cover the whole resonance region
($W\ltap$ 2~GeV) and higher $W$ region.
An essential ingredient to develop a neutrino-nucleus reaction model is
an elementary neutrino-nucleon reaction model, and we will develop one by
extending the DCC model.

\begin{figure}[h]
  \centerline{\includegraphics[width=.35\textwidth]{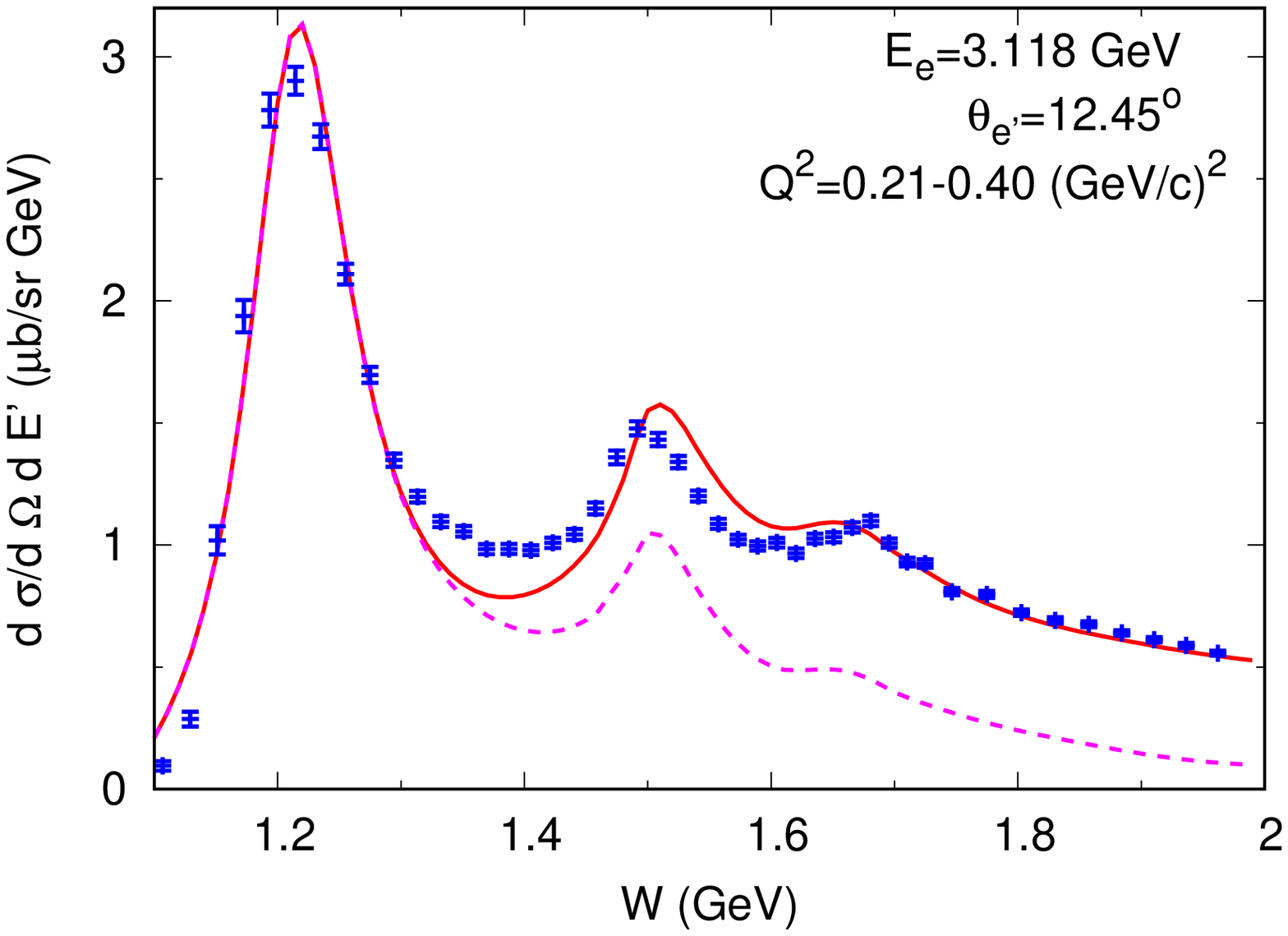}
 \includegraphics[width=.32\textwidth]{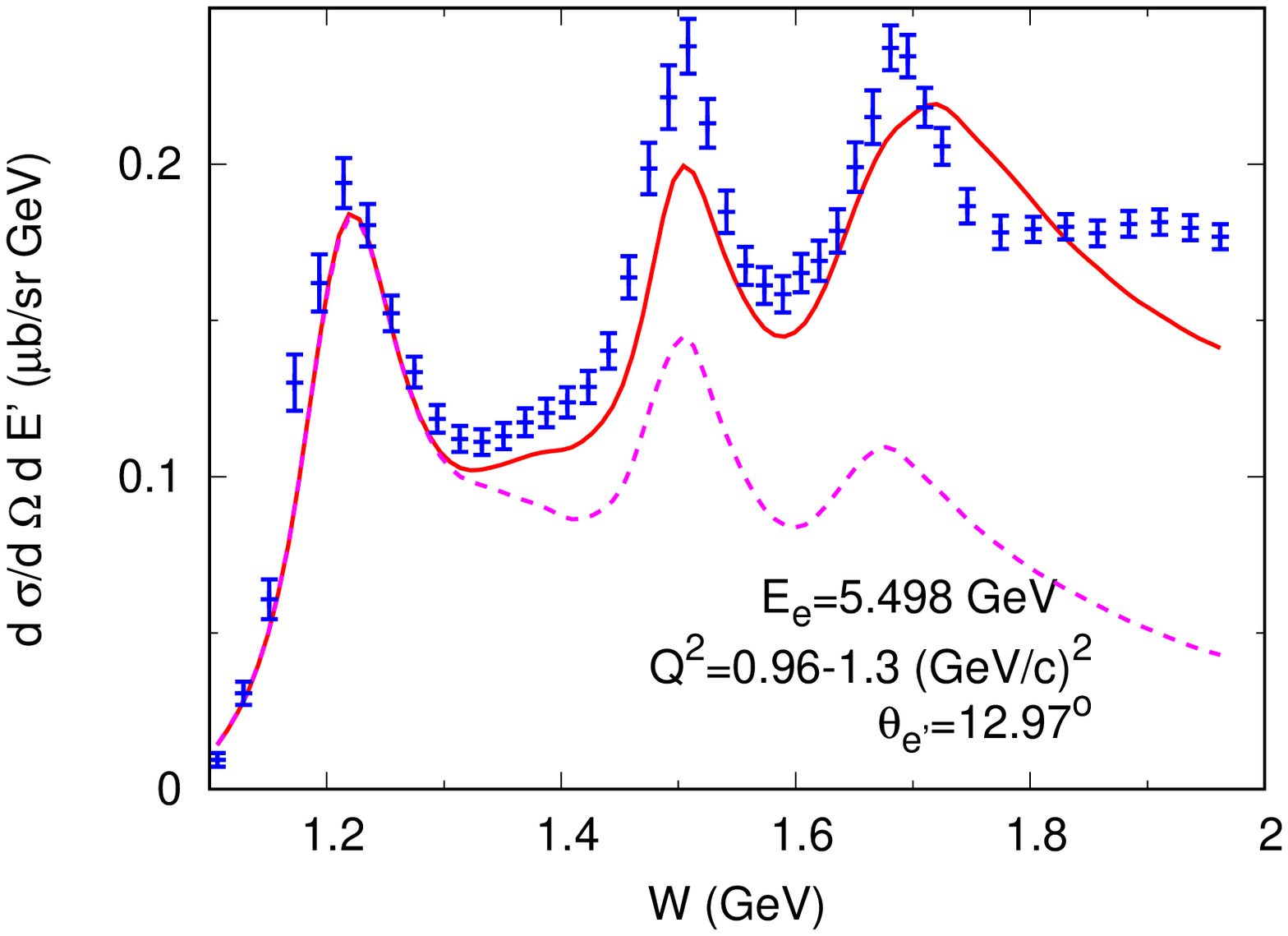}
 \includegraphics[width=.32\textwidth]{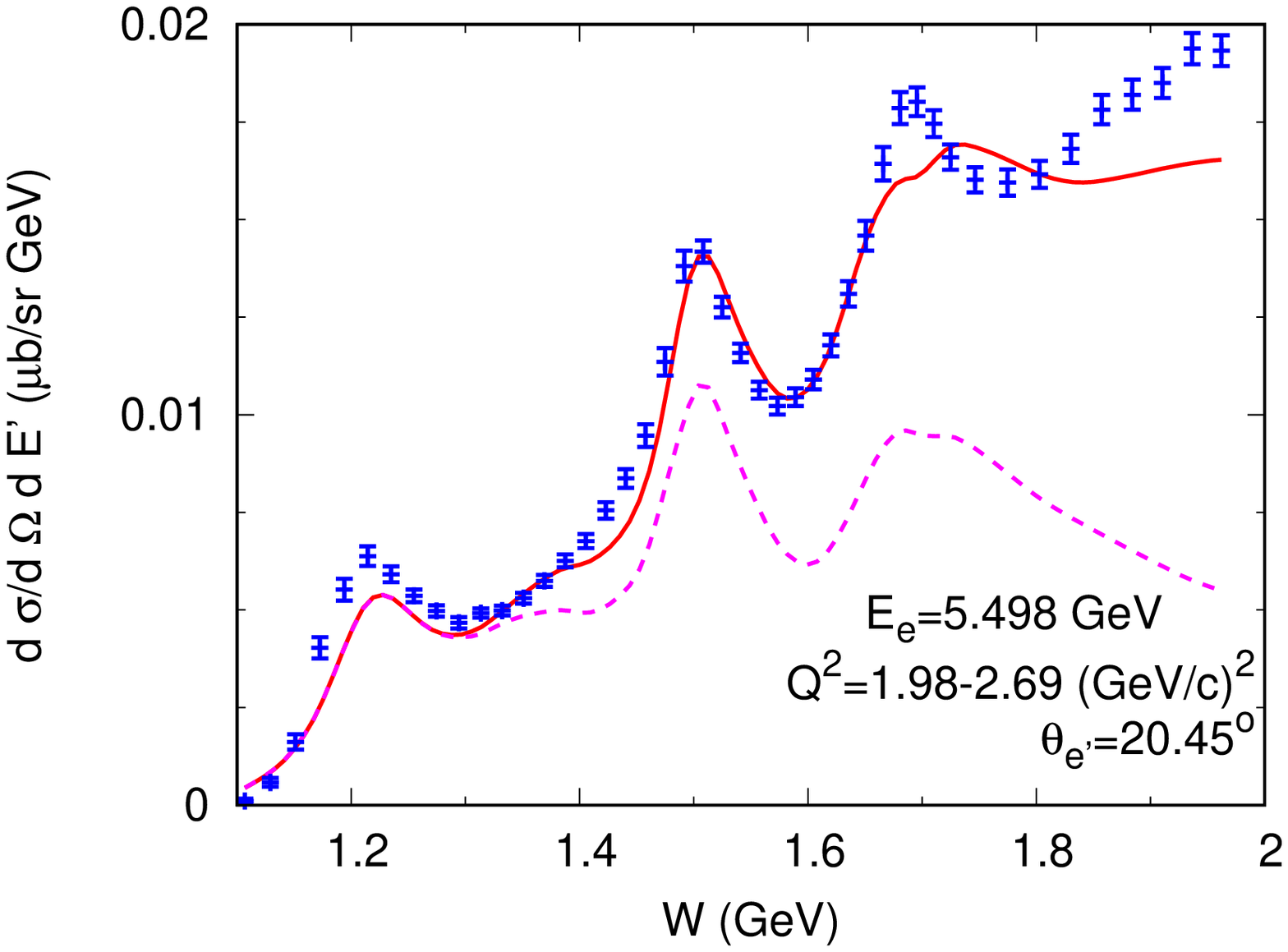}}
 \label{fig:ep-incl}
 \caption{Differential cross sections for inclusive electron-proton scattering
 from the DCC model. 
 The red solid (magenta dashed) curves are for inclusive cross sections
(contributions from the $\pi N$ final states).
The data are from Ref.~\cite{ep-incl-data}.
 }
\end{figure}

First we need to extend the vector current, which has been determined by
analyzing the photo-reaction data, to finite $Q^2$ region.
This amounts to determine the $Q^2$-dependence of the vector $N\to N^*$
transition form factors. 
This can be done by analyzing a good amount of available data
for electron-induced reactions on the nucleon, including both 
single pion productions and inclusive processes.
We analyze both proton- and neutron-target data because,
once the analysis is completed, we need to separate the vector current into
the isovector and isoscalar parts;
the isovector current is necessary to describe charged-current
reactions.
An analysis result for the inclusive electron-proton scattering is shown
in Fig.~\ref{fig:ep-incl}.
The quality of the fit is reasonable enough for an application to the neutrino
reactions.
The two-pion productions give a sizable contribution in the higher $W$
region, as shown by
the difference between the red solid and magenta dashed curves.

Regarding the axial current, we have to take a different strategy
because we have scarce neutrino data available, except for some
useful data for the $\Delta(1232)$ region only.
Thus we follow a guiding principle to derive the axial current:
the PCAC relation with the $\pi N$ reaction amplitudes.
Using the relation, we can relate the axial $N\to N^*$ transition form
factors ($g_{AN\to N^*}$)
at $Q^2\sim 0$ to the $\pi N\to N^*$ coupling strengths
($g_{\pi N\to N^*}$) as
$g_{AN\to N^*}(Q^2\sim 0)\simeq g_{\pi N\to N^*}$;
the phases of the form factors are also fixed.
This can be done only when both $\pi N$ amplitudes and axial currents
are developed consistently with the PCAC relation.
So far, only the DCC model has achieved this~\cite{dcc-nu}.
On the other hand, what has been commonly done is to take 
$N^* \to\pi N$ decay width (from the PDG)
and therefore
$|g_{\pi N\to N^*}|$, and use the relation
$g_{AN\to N^*}(Q^2\sim 0)\simeq |g_{\pi N\to N^*}|$.
Obviously, the phase cannot be determined in a controllable manner.
Regarding the $Q^2$-dependence, we still need an assumption because of
lack of the data. A conventional choice is to use a dipole form factor
with the cutoff of $\sim 1$~GeV.

\begin{figure}[h]
  \centerline{\includegraphics[width=1\textwidth]{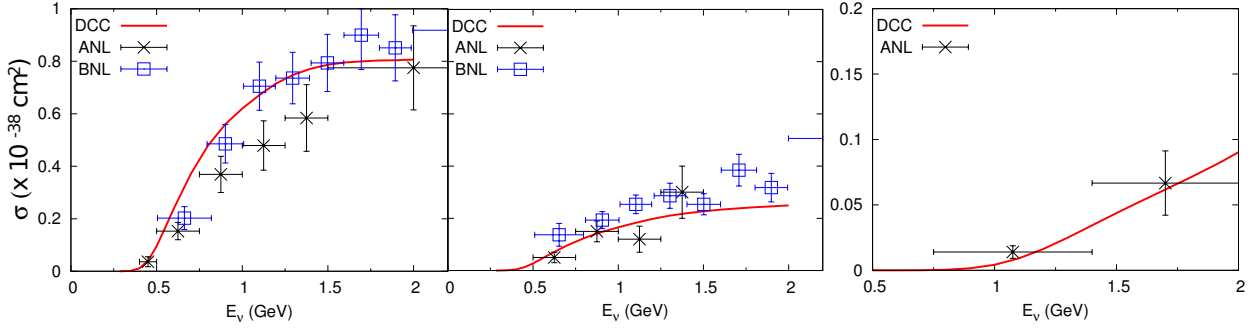}}
 \label{fig:neutrino}
 \caption{Total cross sections for neutrino-nucleon pion productions
 from the DCC model.
 (left) $\nu_\mu p\to\mu^-\pi^+ p$;
 (center) $\nu_\mu n\to\mu^-\pi^0 p$;
 ANL (BNL) data are from Ref.~\cite{anl} (\cite{bnl}). 
 (right) $\nu_\mu p\to\mu^-\pi^+\pi^0 p$ and the data are from
 Ref.~\cite{anl2}.
  Figures taken from Ref.~\cite{dcc-nu}. Copyright (2015) APS.
 }
\end{figure}

With the model setup described above, we make a prediction for
total cross sections of neutrino-nucleon reactions, as shown in 
Fig.~\ref{fig:neutrino}.
The DCC model prediction is consistent with the BNL data.
The model still has a flexibility to adjust the axial $N\to N^*$ form
factors to fit the ANL data.
The DCC model is currently only available neutrino-induced two pion
production model covering the whole resonance region.
The amplitudes from the DCC model are being implemented into ongoing
and forthcoming neutrino oscillation experimental analyses. 

\section{Application of the DCC model to electroweak meson productions
 on deuteron}

\begin{figure}[h]
  \centerline{\includegraphics[width=.85\textwidth]{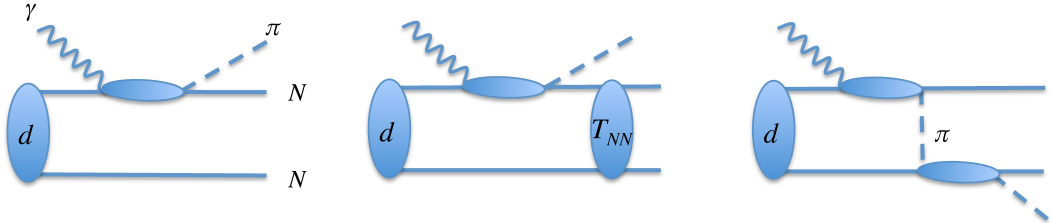}}
 \label{fig:deuteron}
  \caption{The $\gamma d\to \pi NN$ model in this work.
(left) impulse, (center) $NN$ rescattering, (right) $\pi N$ rescattering mechanisms.
 }
 \end{figure}

Our model for electroweak meson productions on the deuteron is based on
 the multiple scattering theory truncated at the first order
 rescattering. 
Therefore our deuteron reaction model consists of the impulse, $NN$
 rescattering, and meson-nucleon rescattering mechanisms, as depicted in
 Fig.~\ref{fig:deuteron}.
 The diagrams are built with elementary (off-shell) amplitudes such as the
 vector and axial currents and meson-baryon scattering
 amplitudes from the DCC model.
 Also, the model includes the $NN$ scattering amplitudes and the
 deuteron wave function for which we employ the CD-Bonn potential~\cite{cdbonn}.

\begin{figure}[h]
  \centerline{\includegraphics[width=1\textwidth]{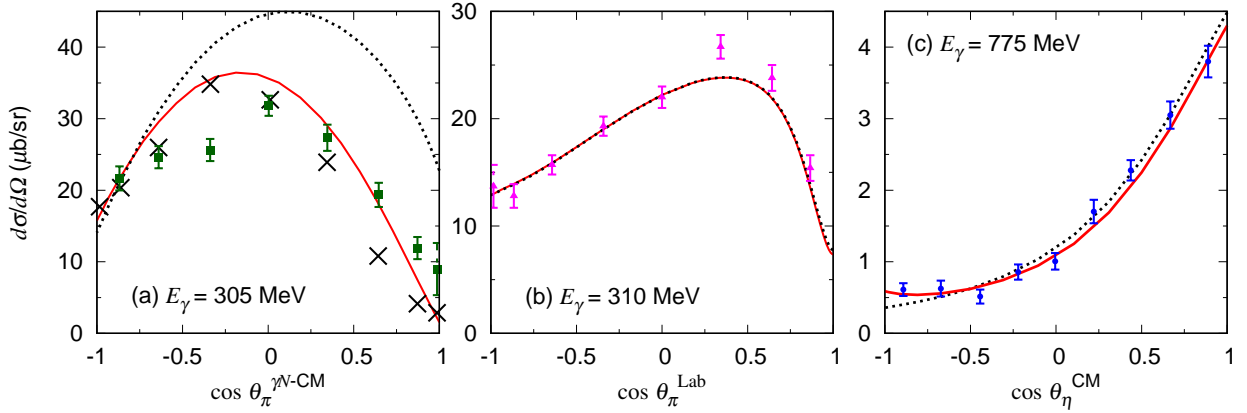}}
 \label{fig:xs-d}
  \caption{ Differential cross sections for meson photoproductions off
  the deuteron.
  (a) $\gamma d\to\pi^0pn$, 
  (b) $\gamma d\to\pi^-pp$, and  
  (c) $\gamma d\to\eta pn$.
  The black dotted curves are calculated with the impulse mechanism
  only, while the red solid curves are from the full model including
  also the $NN$ and meson-nucleon rescattering mechanisms.
  Data are from
  Refs.~\cite{pi0-data1} (green squares), \cite{pi0-data2} (black crosses),
  \cite{pim-data} (magenta triangles), and \cite{eta-data} (blue circles).
  Figures (a,b) taken from Ref.~\cite{dcc-gd} and
  figure (c) from Ref.~\cite{dcc-eta}. Copyright (2017) APS.
 }
 \end{figure}

  Using the model described above,
 we can make a parameter-free prediction for meson photoproduction off
 the deuteron as shown in Fig.~\ref{fig:xs-d}.
 The black dotted curves in the figures are obtained by including the
 impulse mechanism only.
 As seen in Fig.~\ref{fig:xs-d}(a), $\gamma d\to \pi^0pn$ cross sections
 from the impulse approximation significantly overshoot the data.
 However, the final state interactions (FSI) bring the calculation into a good
 agreement with the data. This large FSI effect is due to the
 orthogonality of the $NN$ scattering wave function and the deuteron
 wave function.
 On the other hand, the FSI effect is very small for 
 $\gamma d\to \pi^-pp$ as seen in Fig.~\ref{fig:xs-d}(b).
 Regarding $\gamma d\to \eta pn$ as shown in Fig.~\ref{fig:xs-d}(c),
 the FSI effect does not seem very large, except for the backward $\eta$
 production where a significant enhancement due to $\eta N\to\eta N$
 brings the calculation into  an excellent agreement with the data. 
In what follows, we will study three interesting problems using this DCC-based
deuteron reaction model.

\section{Extraction of neutron-target observables from 
$\gamma d\to \pi NN$}

Data for pion photoproduction off the proton and neutron
constitute a base for studying the baryon spectroscopy.
Because of unavailability of the free neutron target,
the deuteron is the primary target to measure the neutron-target
observables, and we need to understand how to extract it from the
deuteron-target data.
Commonly,
one extracts the $\gamma$-$n$ cross sections by
applying a certain set of kinematical cuts,
assuming that the selected events are from
single-nucleon quasi-free processes.
However, a concern remains whether FSI effects and/or the kinematical
cuts could distort the extracted observables from the free ones.
We will address this question.

Our procedure is as follows.
Starting with the DCC $\gamma N\to \pi N$ amplitudes,
we implement them in the deuteron reaction model and calculate
deuteron-target cross sections to which kinematical cuts are applied.
Using the extraction formula given in Ref.~\cite{prc-extraction},
the neutron-target observables are obtained.
Then the extracted observables are compared with the corresponding free ones
 directly calculated from the DCC $\gamma n\to \pi N$ amplitudes.
 In this analysis, we use realistic kinematical cuts used in
 recent JLab analyses~\cite{jlab1,clas1,clas2}.
 In addition, we also consider a cut on $W$ from the final pion-nucleon kinematics,
 as has been done in the MAMI analysis~\cite{mami1,mami6}.
 In the JLab analyses~\cite{jlab1,clas1,clas2}, on the other hand,
 $W$ is inferred assuming the kinematics where
 the initial neutron is at rest.
 We critically examine the validity of this assumption.

 \begin{figure}[h]
  \centerline{\includegraphics[width=.9\textwidth]{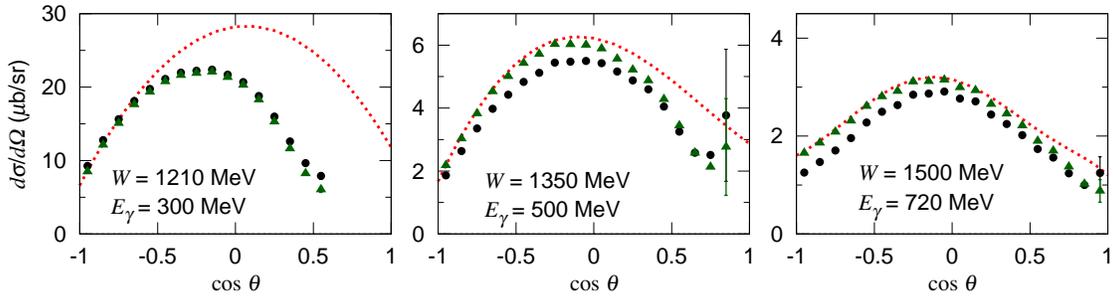}}
 \label{fig:gn-pi0n}
  \caption{Pion angular distribution for $\gamma n\to \pi^0 n$.
The black circles (green triangles) are extracted from $\gamma d\to \pi^0pn$
generated by the DCC-based model including the impulse +
  $NN$-rescattering + $\pi N$-rescattering (impulse +
  $NN$-rescattering) terms.
  The $W$-cut is considered in the extraction. 
  The red dotted curves are the free  $\gamma n\to \pi^0 n$
  cross sections at $W$ from the ANL-Osaka model.
  Figures taken from Ref.~\cite{prc-extraction}. Copyright (2018) APS.
  }
 \end{figure}
The $\gamma n\to\pi^0 n$ differential cross
sections extracted from $\gamma d\to \pi^0 pn$ 
are shown in Fig.~\ref{fig:gn-pi0n}.
The $NN$-rescattering largely reduce the cross sections 
at $E_\gamma = 300$ MeV.
Meanwhile, the $\pi N$-rescattering effect is negligibly small at
$E_\gamma = 300$ MeV, as indicated by 
the small differences between 
the black circles and green triangles.
As the photon energy increases, however, 
the $\pi N$-rescattering becomes comparable to
the $NN$-rescattering, and
significantly reduces the cross sections overall except the forward
pion angles.
Clearly, the kinematical cuts cannot remove the FSI effects, and thus
FSI corrections are necessary.
Also, the FSI effects seen in Fig.~\ref{fig:gn-pi0n}
are qualitatively very similar to, and thus the first theoretical
explanation of, those found in 
the MAMI analysis~\cite{mami1,mami6}.

 \begin{figure}[h]
  \centerline{\includegraphics[width=1\textwidth]{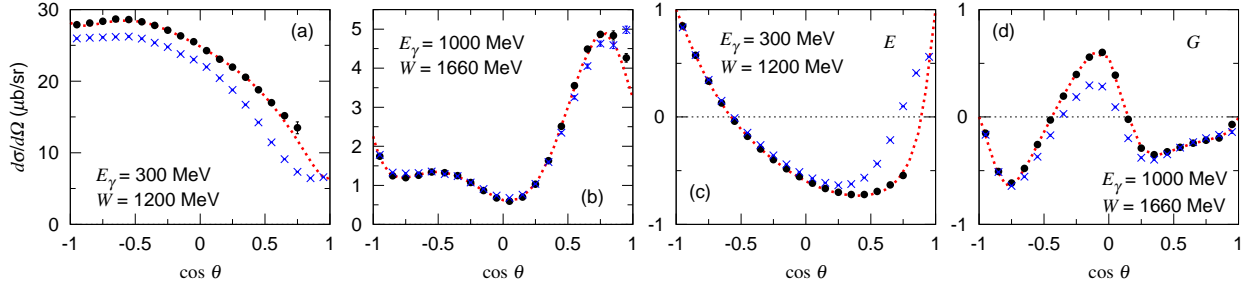}}
 \label{fig:gn-pimp-qf}
  \caption{
  (a,b) Unpolarized differential cross sections for $\gamma n\to \pi^-p$
 extracted from $\gamma d\to \pi^-pp$
generated from the DCC-based model including only the quasi-free
  mechanism.
The black circles [blue crosses] are
  extracted with [without] $W$ cut.
  (c)[(d)] The polarization observables $E$ [$G$]
   for $\gamma n\to \pi^-p$
 extracted from $\gamma d\to \pi^-pp$.
  The other features are the same as those in Fig.~\ref{fig:gn-pi0n}.
  Figures taken from Ref.~\cite{prc-extraction}. Copyright (2018) APS.
  }
 \end{figure}
Now we examine the extraction without the $W$-cut as employed in the
recent JLab analyses.
For this study, we do not consider the FSI.
The $\gamma n\to\pi^-p$ observables including the polarization observables of
$E$ and $G$ are shown in Fig.~\ref{fig:gn-pimp-qf}.
With the $W$-cut, the extracted observables (black circles) accurately reproduce the
free ones (red dotted curves).
Without the $W$-cut, however, 
 the extracted observables (blue crosses) sometimes significantly
 deviate from the free ones.
 This deviation is caused by the Fermi motion.
 The result indicates that it is important to apply the $W$-cut to
 suppress the problematic Fermi motion effect, thereby extracting the
 neutron-target observables accurately.

\section{Low-energy $\eta$-nucleon interaction studied with
$\gamma d\to \eta np$}

The $\eta$-nucleon scattering length ($a_{\eta N}$) governs the low-energy behavior of
 the $\eta$-nucleon scattering which is a basic feature of the meson-baryon dynamics.
 Also, the existence of exotic $\eta$-mesic nuclei strongly depends on 
its value~\cite{etan9}.
Being the important quantity, however,
$a_{\eta N}$ has not been well determined yet. 
Several coupled-channel analyses have been done on 
the $\pi N\to \pi N, \eta N$ and $\gamma N\to \pi N, \eta N$
reaction data to determine $a_{\eta N}$.
The $pn \to \eta d$ reaction has also been analyzed.
The imaginary part of $a_{\eta N}$ 
from these analyses is fairly consistent, 
falling into ${\rm Im} [a_{\eta N}]=0.2$--0.3~fm.
However, the real part is in a significantly wider range of
${\rm Re} [a_{\eta N}]=0.2$--0.9~fm~\cite{etan9}.
Because $a_{\eta N}$ has been extracted from the indirect information,
the model dependence is difficult to avoid.
To better determine $a_{\eta N}$, we need a process that sensitively
probes the $\eta N\to\eta N$ scattering, while the other background mechanisms
being suppressed.

A realization of this idea is the ongoing experiment
at the Research Center
for Electron Photon Science (ELPH), Tohoku University~\cite{plan}.
In this experiment,
the $\gamma d\to \eta pn$ cross section is measured at a special kinematics:
$E_\gamma\sim 0.94$~GeV and $\theta_p\sim 0^\circ$
($\theta_p$: angle between the scattered proton and the incident photon).
At this kinematics, 
the produced $\eta$ is almost at rest and very likely to 
interact with the spectator neutron.
The scattered proton with a large momentum has
little chance to interact with the $\eta$ and neutron.
We refer to this special kinematics as the ELPH kinematics.
The ELPH kinematics seems ideal to study the low-energy $\eta$-nucleon
interaction.
A model is still necessary to extract 
$a_{\eta N}$ from the ELPH data, and the DCC-based
$\gamma d\to \eta pn$ model is a promising option.
Using this model, in this work, we study 
$\gamma d\to \eta pn$ at the ELPH kinematics,
and examine the sensitivity of the ELPH experiment to 
$a_{\eta N}$.

 \begin{figure}[h]
  \centerline{
  \includegraphics[width=.5\textwidth]{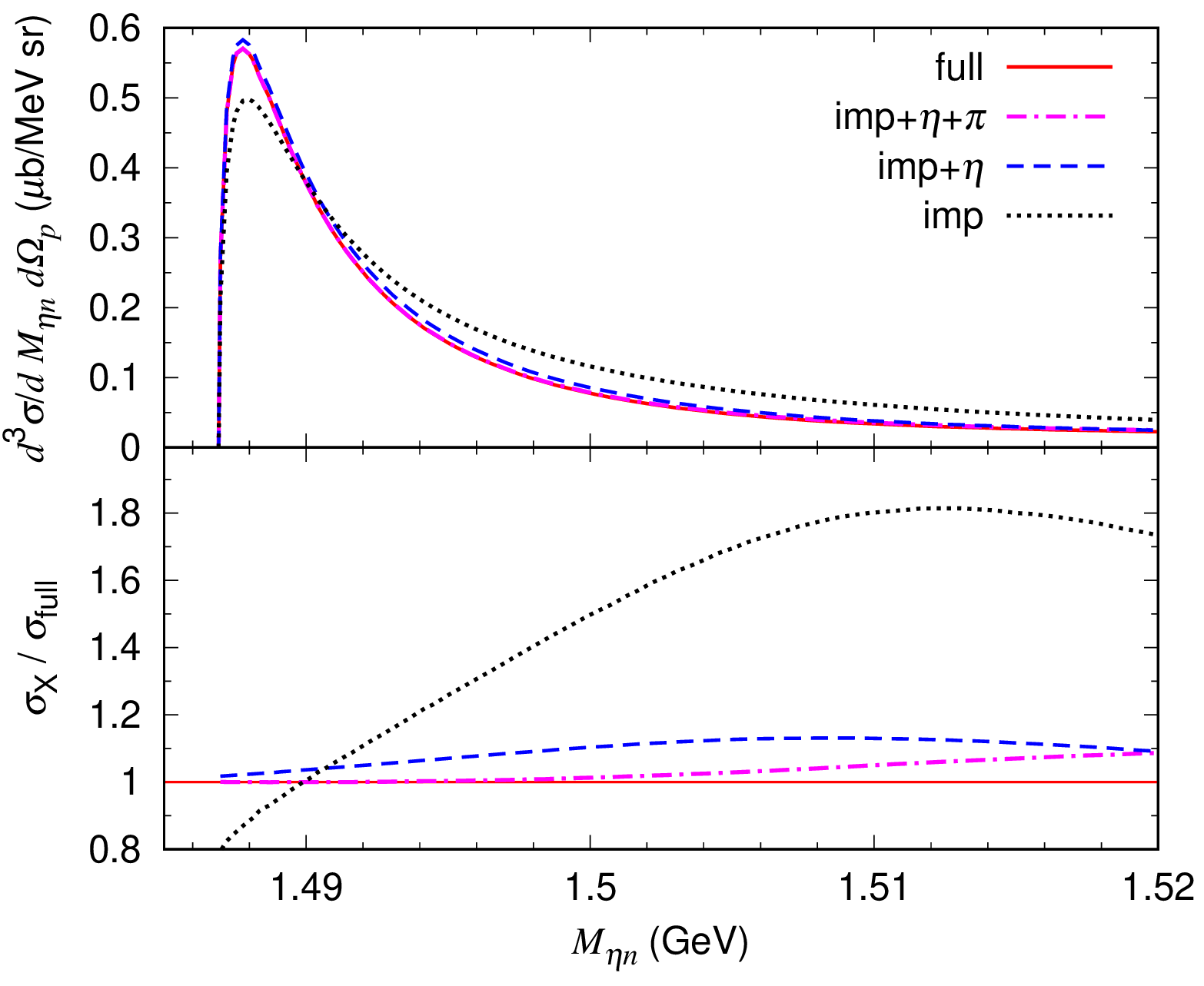}
  \includegraphics[width=.5\textwidth]{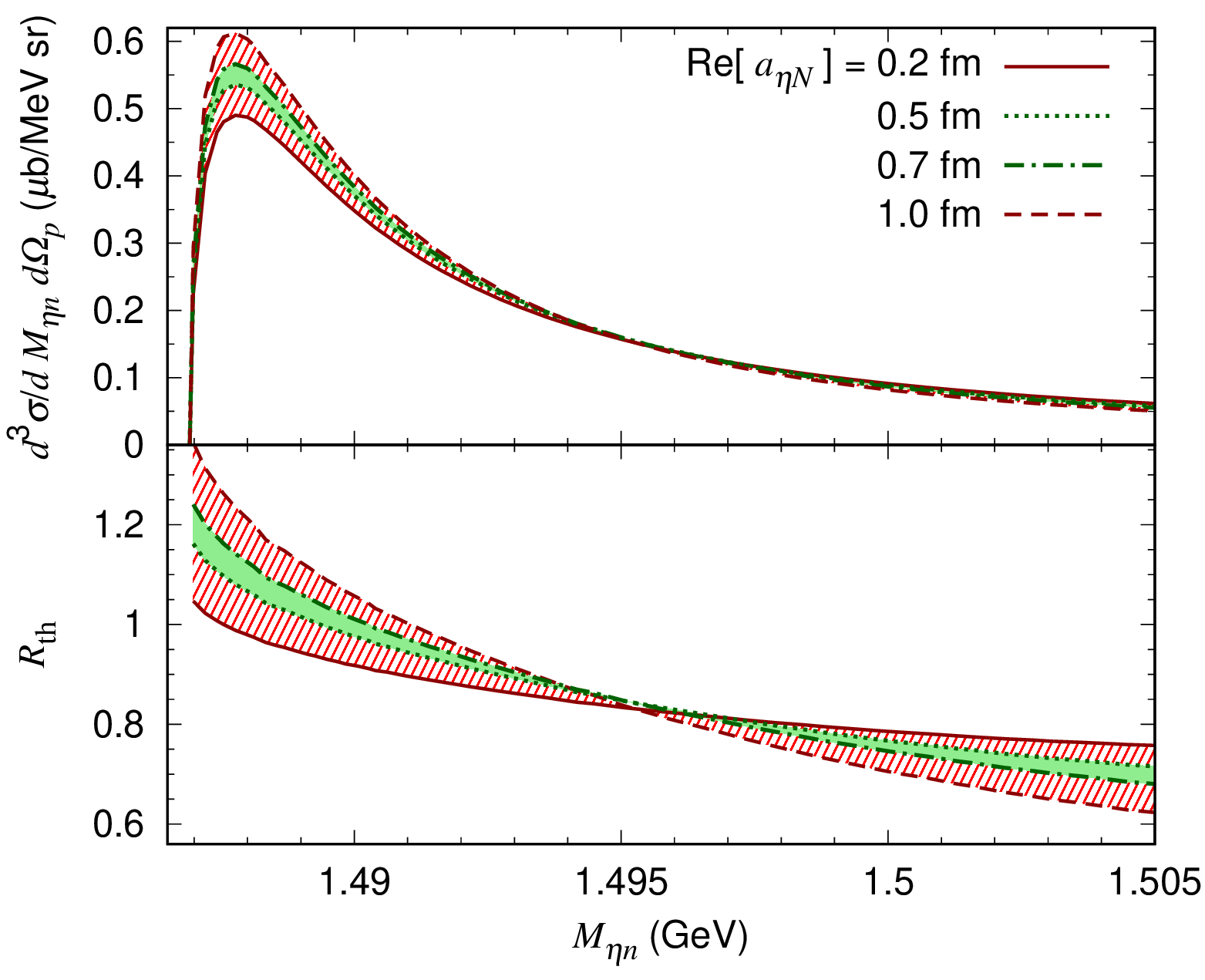}
  }
 \label{fig:gd-eta}
  \caption{
(Left,Top) Differential cross section for
$\gamma d\to \eta pn$ at $E_\gamma=0.94$~GeV and $\theta_p=0^\circ$.
  The solid curve is from the full calculation while the dotted curve includes
  the impulse mechanism only.
  The dashed curve includes 
  the impulse and $\eta$-exchange mechanisms, 
  and the dash-dotted curve additionally includes 
  the pion-exchange.
  The dash-dotted curve overlaps almost exactly with the solid curve.
  (Left,Bottom) The differential
cross sections calculated with the various mechanisms divided by the
full calculation.
 (Right,Top) Re[$a_{\eta N}$]-dependence of $\gamma d\to\eta pn$ differential
  cross sections at $E_\gamma=0.94$~GeV and $\theta_p=0^\circ$.
  The curves are obtained with the full model by varying
  Re[$a_{\eta N}$] = 0.2,
  0.5, 0.7, and 1.0~fm; Im[$a_{\eta N}$] = 0.25~fm and $r_{\eta N}$ = 0.
  (Right,Bottom) The ratio $R_{\rm th}$ of Eq.~(\ref{eq:Ratio}) for various values
of Re[$a_{\eta N}$].
    Figures taken from Ref.~\cite{dcc-eta}. Copyright (2017) APS.
}
 \end{figure}
 In Fig.~\ref{fig:gd-eta}(left,top),
differential cross sections for 
$\gamma d\to \eta pn$ at the ELPH kinematics
is shown as a function of the $\eta$-neutron invariant mass ($M_{\eta n}$).
While the impulse mechanism is dominant, 
the $\eta n\to \eta n$ FSI mechanism gives a sizable effect on 
the cross section:  $-$40 to +20\%.
On the other hand, as we expected for the ELPH kinematics, 
the other FSI mechanisms are well suppressed.
The $\pi n\to\eta n$ FSI, for which we have data and thus
control well, can change the cross sections
by $\ltap$9\%, and the $NN$ rescattering effect
is very small for $M_{\eta n}\ltap 1.5$~GeV.
The result indicates that the proton is well isolated from
the $\eta n$ system, and thus we can safely neglect
multiple rescatterings beyond the first-order rescattering.

Having shown that $\gamma d\to \eta pn$ cross sections at the ELPH
kinematics are largely influenced by 
the $\eta n\to \eta n$ FSI,
the next question is how sensitive the ELPH data is to $a_{\eta N}$.
To address this question, in the $\gamma d\to \eta pn$ model,
we replace the DCC $\eta n\to \eta n$ amplitudes with those of the
effective range expansion.
We then vary $a_{\eta N}$ and $r_{\eta N}$ (effective range)
to examine how sensitively 
the $\gamma d\to \eta pn$ cross sections change.
The result is shown in Fig.~\ref{fig:gd-eta}(right,top) where
${\rm Re}[a_{\eta N}]$ is varied.
The obtained cross sections are mostly within the red striped region.
The change of the cross sections is more clearly seen
in Fig.~\ref{fig:gd-eta}(right,bottom) showing 
the ratio defined by
\begin{eqnarray}
\label{eq:Ratio}
R_{\rm th}(M_{\eta n}) = {d^3\sigma_{\rm full}/dM_{\eta n} d\Omega_p |_{\theta_p=0^\circ}
\over d^3 \sigma_{\rm imp}/dM_{\eta n} d\Omega_p|_{\theta_p=0^\circ}} 
\ ,
\end{eqnarray}
where $\sigma_{\rm full}$ ($\sigma_{\rm imp}$) is the cross section from
the full model (impulse approximation).
Because the ELPH are measuring both the proton and deuteron target data,
and thus the experimental counterpart to $R_{\rm th}$ will be available.
As shown in the figures,
$R_{\rm th}$ changes by $\sim$25\% within the red striped region
at $M_{\eta n}\sim 1.488$~GeV 
of the quasi-free (QF) peak.
When ${\rm Re}[a_{\eta N}]$ is varied by $\pm 0.1$~fm from 0.6~fm,
meanwhile,
the cross sections change by $\sim$5\% at the QF peak
as shown by the green solid bands.
This indicates that 
$R_{\rm expt}$ data of 5\% error per MeV bin
can determine ${\rm Re}[a_{\eta N}]$ 
at the precision of $\sim\pm 0.1$~fm.
This is a significant improvement over
the current uncertainty.
The ELPH experiment is capable of achieving
this precision measurement.

 \section{FSI corrections to neutrino-nucleon cross section data from
 neutrino-deuteron experiments}

A reliable neutrino-nucleon reaction model is
a key ingredient in developing a neutrino-nucleus reaction model
to be used in neutrino-oscillation analyses.
Regarding the single pion productions ($\nu N\to l\pi N$),
many microscopic models with different dynamical contents
have been developed.
An overview of these microscopic models can be found in 
Ref.~\cite{nu-review}, and a detailed comparison in Ref.~\cite{pi_comp}.
We stress that 
the total cross section data~\cite{anl,bnl}
of $\nu_\mu p\to \mu^-\pi^+ p$,
$\nu_\mu n\to \mu^-\pi^+ n$, and $\nu_\mu n\to \mu^-\pi^0 p$
play a crucial role in developing these models.
All the models include the axial $\Delta(1232)$-excitation mechanism,
and the strength of this dominant piece is always fitted to
the total cross section data.
However, the total cross section data currently available were actually
extracted from neutrino-deuteron reaction ($\nu_\mu d\to \mu^-\pi NN$)
data under an assumption that 
the quasifree mechanism dominates and FSI are negligible.
Considering the precision needed for neutrino-nucleus reaction
models in near-future neutrino oscillation analyses,
we can no longer ignore the possible FSI effects, and thus we address
this problem.

 \begin{figure}[h]
  \centerline{
\includegraphics[width=.9\textwidth]{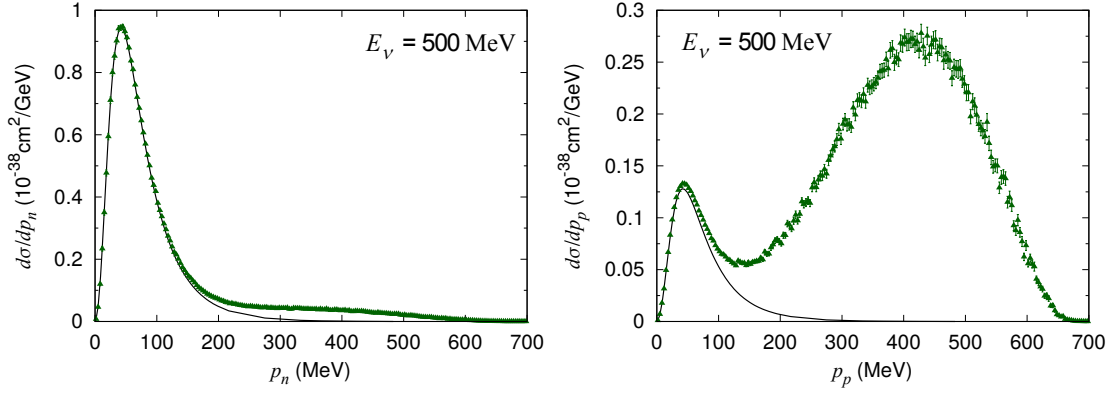}
  }
 \label{fig:spec}
  \caption{Neutron (left) and proton (right) momentum distributions in
  $\nu_\mu d\to\mu^-\pi^+pn$ at $E_\nu=0.5$~GeV.
  The impulse mechanism gives the green triangles.
  The black solid curve in the left (right) panel
  is obtained by convoluting 
 the $\nu_\mu p\to\mu^-\pi^+p$ ($\nu_\mu n\to\mu^-\pi^+n$)
 cross sections with the deuteron wave function as in 
Eq.~(\ref{eq:conv}).
  The error bars are statistical from using the Monte-Carlo method for the
  phase-space integral.
  }
 \end{figure}
We analyze the spectator momentum ($p_s$) distribution in
$\nu_\mu d\to\mu^-\pi N'N_s$ ($N_s$: spectator).
The $p_s$-distribution is a minimal information to extract
the cross section for
$\nu_\mu p\to\mu^-\pi^+ p$ ($\nu_\mu n\to\mu^-\pi^+ n$).
This is because the quasifree
neutrino-proton (neutrino-neutron) pion production process
is expected to dominate exclusively in a low-$p_n$ ($p_p$) region.
This can been seen in Fig.~\ref{fig:spec} where 
we show the neutron and proton momentum distributions
calculated with the impulse approximation,
along with the $\nu_\mu N\to\mu^-\pi N'(\equiv\alpha)$ cross section
convoluted with the deuteron wave function ($\Psi_d$):
\begin{eqnarray}
 {d\tilde\sigma_{\alpha}(E_\nu)\over dp_s} = p_s^2 \int d\Omega_{p_s}
  \sigma_{\alpha} (\tilde E_{\nu}) |\Psi_d(\vec p_s)|^2
  \ ,
\label{eq:conv}
\end{eqnarray}
where the total cross section $\sigma_{\alpha}$ is calculated
with the same $\nu_\mu N\to\mu^-\pi N'$ amplitudes implemented in the 
$\nu_\mu d\to\mu^-\pi NN$ model; $\tilde E_{\nu}$ is
the boosted neutrino energy.
As $p_s$ increases, the convoluted cross sections undershoot the impulse
calculation because
the other nucleon's contribution becomes more significant. 
This is more evident in the $p_p$-distribution in 
Fig.~\ref{fig:spec}(right)
because the cross section of $\nu_\mu p\to\mu^-\pi^+ p$ 
is $\sim 9$ times larger than that of $\nu_\mu n\to\mu^-\pi^+ n$ 
at this neutrino energy.

 \begin{figure}[h]
  \centerline{
\includegraphics[width=.95\textwidth]{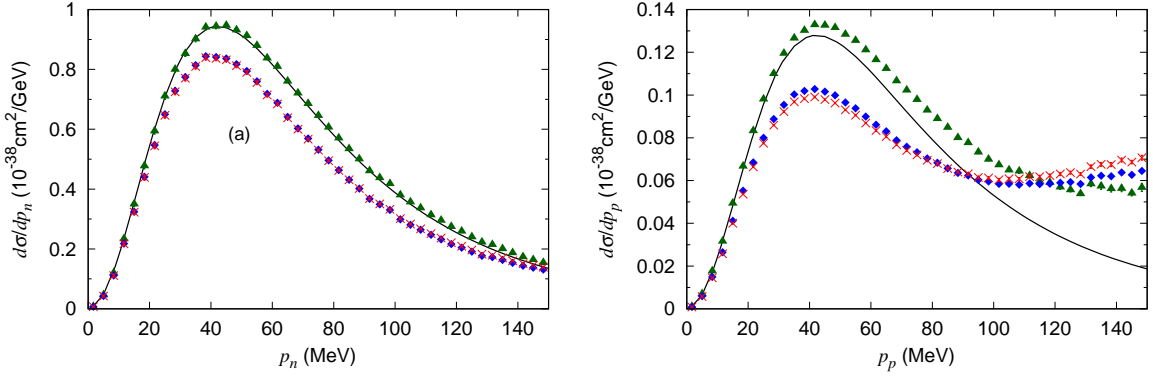}
  }
 \label{fig:gd-nud}
  \caption{Neutron (left) and proton (right) momentum distributions in
  $\nu_\mu d\to\mu^-\pi^+pn$ at $E_\nu=0.5$~GeV.
  The impulse mechanism gives the green triangles, while
  the $NN$ rescattering mechanism is also included in the blue diamonds.
  The red crosses are calculated with the full model including
the impulse + $NN$ + $\pi N$ rescattering mechanisms.
 The $\nu_\mu p\to\mu^-\pi^+p$ and $\nu_\mu n\to\mu^-\pi^+n$
 cross sections are convoluted with the deuteron wave function to give 
  the black solid curves in the left and right panels, respectively.
  Figures taken from Ref.~\cite{dcc-nud}. Copyright (2019) APS.
  }
 \end{figure}
 Now we examine the FSI effects in Fig.~\ref{fig:gd-nud}.
 The $p_s$-distribution ($d\sigma_{\nu_\mu d}/dp_s$) for
$\nu_\mu d\to\mu^-\pi^+ pn$ is reduced significantly by
the $NN$ FSI as seen in  the differences between 
the blue diamonds and green triangles in the figures.
In particular,  the quasifree peak in the low-$p_s$ region is
 significantly lowered.

  \begin{figure}[h]
  \centerline{
\includegraphics[width=1\textwidth]{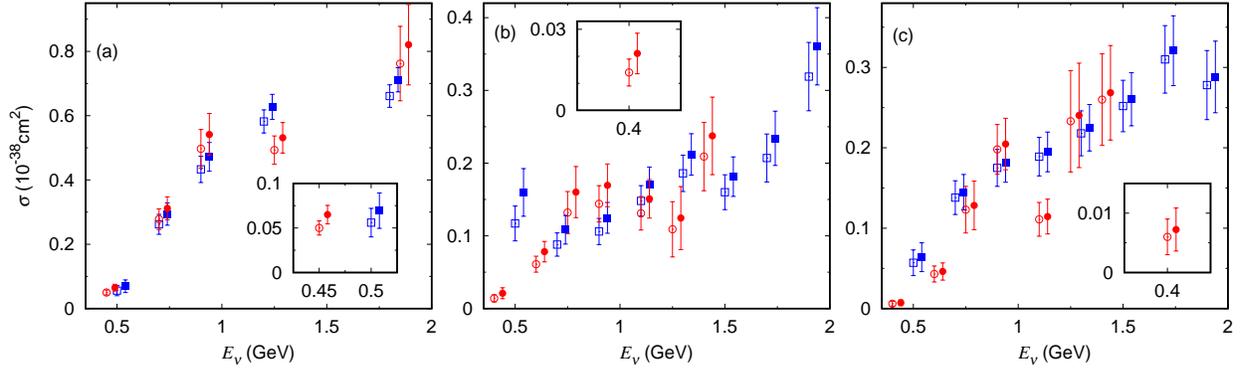}
  }
 \label{fig:nu-corrected}
   \caption{Total cross sections for
   (a) $\nu_\mu p\to\mu^-\pi^+p$,
   (b) $\nu_\mu n\to\mu^-\pi^+n$, and
   (c) $\nu_\mu n\to\mu^-\pi^0p$.
The open red circles and open blue squares are the reanalyzed ANL and BNL
   data~\cite{reanalysis}, respectively, without applying cuts on $W$.
The ANL and BNL data are corrected for the Fermi motion and FSI and
   shown by the solid red circles and solid blue squares, respectively.
   The inlet shows an enlargement of the small $E_\nu$ region.
    Figures taken from Ref.~\cite{dcc-nud}. Copyright (2019) APS.
  }
 \end{figure}
 The significant FSI effects found above
 points to the necessity of correcting
the deuterium bubble chamber data for the
$\nu_\mu N\to\mu^-\pi N$ total cross sections~\cite{anl,bnl}.
Using the FSI effects shown in Fig.~\ref{fig:gd-nud},
we correct the data and show them in 
Fig.~\ref{fig:nu-corrected} along with 
the original ones~\cite{reanalysis} for
comparison.
The correction
enhances the cross sections by factors of
1.05--1.12, 1.10--1.27, and 1.01--1.02 
for 
$\nu_\mu p\to\mu^-\pi^+p$, $\nu_\mu n\to\mu^-\pi^+n$, and
$\nu_\mu n\to\mu^-\pi^0p$, respectively.
The correction is larger for smaller $E_\nu$.

\section{Summary}

I reviewed our recent activity with the DCC approach.
I discussed the DCC approach to the single nucleon sector
such as the DCC analysis of 
$\pi N, \gamma N\to \pi N, \pi\pi N, \eta N, K\Lambda, K\Sigma$
reactions, and its extension to finite $Q^2$ region and
neutrino-induced meson productions.
Then I discussed applications of the DCC model amplitudes to
electroweak meson productions on the deuteron:
(i) the extraction of neutron-target observables from 
$\gamma d\to \pi NN$;
(ii) a novel method of extracting 
$\eta$-nucleon scattering length from
$\gamma d\to \eta np$;
(iii) FSI corrections to neutrino-nucleon cross section data from
neutrino-deuteron experiments.
Although not covered during the presentation because of the time limitation,
I mention that
the DCC approach has also been applied to analyzing $\bar K N$ reactions
to extract hyperon resonance properties~\cite{ystar1,ystar2,ystar3},
and to describing FSI in heavy meson decays into three mesons~\cite{3pi1,3pi2,3pi3}.


\section{ACKNOWLEDGMENTS}
The author thanks H. Kamano, T.-S.H. Lee, T. Sato, and T. Ishikawa for
collaborations on the subjects discussed in this manuscript. 
This work is in part supported by 
National Natural Science Foundation of China (NSFC) under contracts 11625523.


\nocite{*}
\bibliographystyle{aipnum-cp}%

\begin{thebibliography}{}
\bibitem{web}
H.  Kamano,  T.-S. H. Lee, S.X. Nakamura, and T. Sato,
http://www.phy.anl.gov/theory/research/anl-osaka-pwa/

\bibitem{bg12}
	A.V. Anisovich, R. Beck, E. Klempt, V.A. Nikonov,
	A.V. Sarantsev, and U. Thoma, 
Eur. Phys. J. A {\bf 48},15 (2012).
	
\bibitem{knls13}
H.~Kamano, S.~X.~Nakamura, T.-S.~H.~Lee, and T.~Sato,
Phys. Rev. C {\bf 88}, 035209 (2013).

\bibitem{knls16}
H. Kamano, S.X. Nakamura, T.-S.H. Lee, and T. Sato, Phys. Rev. C {\bf 94}, 015201 (2016).

\bibitem{t2k}
http://t2k-experiment.org

\bibitem{dune}
http://www.dunescience.org
	
\bibitem{ep-incl-data}
Preliminary results from JLab E00-002, C. Keppel, M.I. Niculescu,
	spokespersons. Data files can be obtained at	
	https://hallcweb.jlab.org/resdata/database.
	
 \bibitem{dcc-nu}
 S.X. Nakamura, H. Kamano, and T. Sato, Phys. Rev. D {\bf 92}, 074024 (2015).
	 
\bibitem{anl} S. J. Barish et al., Phys. Rev. D {\bf 19}, 2521 (1979).

\bibitem{bnl} T. Kitagaki et al., Phys. Rev. D {\bf 34}, 2554 (1986).

 \bibitem{anl2} D. Day et al., Phys. Rev. D {\bf 28}, 2714 (1983).

\bibitem{cdbonn}
R. Machleidt, 
Phys. Rev. C {\bf 63}, 024001 (2001).

 \bibitem{pi0-data1} B. Krusche et al., Eur. Phys. J. A {\bf 6}, 309 (1999).
 \bibitem{pi0-data2} U. Siodlaczek et al., Eur. Phys. J. A {\bf 10}, 365 (2001).
 \bibitem{pim-data} P. Benz et al. (Aachen-Bonn-Hamburg-Heidelberg-Muenchen Collaboration), Nucl. Phys.
B {\bf 65}, 158 (1973).
 \bibitem{eta-data} B. Krusche et al., Phys. Lett. B {\bf 358}, 40 (1995).

 \bibitem{dcc-gd}
 S.X. Nakamura, H. Kamano, T.-S.H. Lee, and T. Sato, arXiv:1804.04757.

 \bibitem{dcc-eta}
S.X. Nakamura, H. Kamano, and T. Ishikawa, Phys. Rev. C {\bf 96}, 042201(R) (2017).

 \bibitem{prc-extraction}
 S.X. Nakamura, Phys. Rev. C {\bf 98}, 042201(R) (2018).

\bibitem{jlab1}
W.J. Briscoe, A.E. Kudryavtsev, P. Pedroni, I.I. Strakovsky, V.E. Tarasov, and R.L. Workman, 
Phys. Rev. C {\bf 86}, 065207 (2012).

\bibitem{clas1}
D. Ho et al. (CLAS Collaboration), 
Phys. Rev. Lett. {\bf 118}, 242002 (2017).

\bibitem{clas2}
P.T. Mattione et al. (CLAS Collaboration),
Phys. Rev. C {\bf 96}, 035204 (2017).
	
\bibitem{mami1}
M. Dieterle et al. (A2 Collaboration),
Phys. Rev. Lett. {\bf 112}, 142001 (2014).

 \bibitem{mami6}
M. Dieterle et al. (A2 Collaboration),
Phys. Rev. C {\bf 97}, 065205 (2018).

\bibitem{etan9}
Q. Haider and L. C. Liu, Int. J. Mod. Phys. E {\bf 24}, 1530009 (2015).

\bibitem{plan} 
T.~Ishikawa {\it et al.}, JPS Conf.\ Proc.\ {\bf 13}, 020031 (2017).

\bibitem{nu-review}
S.X. Nakamura, H. Kamano, Y. Hayato, M. Hirai, W. Horiuchi, S. Kumano,
	T. Murata, K. Saito, M. Sakuda, T. Sato, and Y. Suzuki, 
Rep. Prog. Phys. {\bf 80}, 056301 (2017).

\bibitem{pi_comp}
J.E. Sobczyk, E. Hern\'andez, S.X. Nakamura, J. Nieves, and T. Sato,
Phys. Rev. D {\bf 98}, 073001 (2018).

 \bibitem{dcc-nud}
S.X. Nakamura, H. Kamano, and T. Sato, Phys. Rev. D {\bf 99}, 031301(R) (2019).

 \bibitem{reanalysis}
P. Rodrigues, C. Wilkinson, and K. McFarland, Eur. Phys. J. C {\bf 76}, 474 (2016).

 \bibitem{ystar1}
H. Kamano, S.X. Nakamura, T.-S.H. Lee, and T. Sato, Phys. Rev. C {\bf 90}, 065204 (2014).

 \bibitem{ystar2}
H. Kamano, S.X. Nakamura, T.-S.H. Lee, and T. Sato, Phys. Rev. C {\bf 92}, 025205 (2015).
	
 \bibitem{ystar3}
H. Kamano and T.-S.H. Lee, Phys. Rev. C {\bf 94}, 065205 (2016).
	 
 \bibitem{3pi1}
H. Kamano, S.X. Nakamura, T.-S.H. Lee, and T. Sato, Phys. Rev. D {\bf 84}, 114019 (2011).

 \bibitem{3pi2}
S.X. Nakamura, H. Kamano, T.-S.H. Lee, and T. Sato, Phys. Rev. D {\bf 86}, 114012 (2012).

 \bibitem{3pi3}
S.X. Nakamura, Phys. Rev. D {\bf 93}, 014005 (2016).
	 
\end{thebibliography}

\end{document}